\documentclass[]{elsart}
\usepackage{pgf}
\usepackage[english]{babel}
\usepackage{subeqn}
\usepackage{epsfig}
\usepackage[english]{babel}
\usepackage{amsfonts}
\usepackage{url}

\setkeys{Gin}{width=5in} % Set global figure size

\begin{document}

%%%%%%%%%%%%%%%%%%%%%%%%%%%%%%%%%%%% Include Files

\begin{frontmatter}

\title{Neutronic Design and Measured Performance of the Low Energy Neutron Source (LENS) Target Moderator
Reflector Assembly}

\author[IUCF]{C.M. Lavelle\corauthref{cor}}
\corauth[cor]{Corresponding author.} \ead{chlavell@indiana.edu}
\author[IUCF]{D.V. Baxter }\ead{baxterd@indiana.edu}
\author[IUCF]{A. Bogdanov}
\author[IUCF]{V.P. Derenchuk}
\author[IUCF]{H. Kaiser}
\author[IUCF]{M.B. Leuschner}
\author[Fresno]{M. A. Lone}
\author[IUCF]{W. Lozowski}
\author[IUCF]{H. Nann}
\author[IUCF]{B. v.Przewoski}
\author[IUCF]{N. Remmes}
\author[IUCF]{T. Rinckel}
\author[IUCF]{Y. Shin}
\author[IUCF]{W.M. Snow}
\author[IUCF]{P.E. Sokol}

\address[IUCF]{IUCF 2401 Milo B. Sampson Lane, Bloomington, IN 47408}
\address[Fresno]{1449 W Barstow Ave, Fresno, CA, USA 93711}

\begin{abstract}
The Low Energy Neutron Source (LENS) is an accelerator-based pulsed
cold neutron facility under construction at the Indiana University
Cyclotron Facility (IUCF). The idea behind LENS is to produce pulsed
cold neutron beams starting with $\sim$MeV neutrons from (p,n)
reactions in Be which are moderated to meV energies and extracted from
a small solid angle for use in neutron instruments which can operate
efficiently with relatively broad ($\sim$1 msec) neutron pulse widths.
Although the combination of the features and operating parameters of
this source is unique at present, the neutronic design possesses
several features similar to those envisioned for future neutron
facilities such as long-pulsed spallation sources (LPSS) and very cold
neutron (VCN) sources. We describe the underlying ideas and design
details of the target/moderator/reflector system (TMR) and compare
measurements of its brightness, energy spectrum, and emission time
distribution under different moderator configurations with MCNP
simulations.  Brightness measurements using an  ambient temperature
water moderator agree with MCNP simulations  within the 20\% accuracy
of the measurement.  The measured neutron emission time distribution
from a solid methane moderator is in agreement with simulation and the
cold neutron flux is sufficient for neutron scattering studies of
materials. We describe some possible modifications to the existing
design which would increase the cold neutron brightness with negligible
effect on the emission time distribution. \footnote{This is a preprint
version of an article which has been published in Nuclear Instruments
and Methods in Physics Research A 587 (2008) 324-341.

\url{http://dx.doi.org/10.1016/j.nima.2007.12.044}}

\end{abstract}


\end{frontmatter}

\section{Introduction}
\label{sec:intro}
The Low Energy Neutron Source (LENS) at Indiana University is a pulsed
neutron source based on (p,xn) reactions in beryllium for proton
energies of 13 MeV or less \cite{LENS2005c,LENS2005d}.  This relatively
small-scale neutron source is designed to provide pulsed cold neutron
beams of sufficient intensity to conduct neutron  research with
relatively low cost and minimal source activation, thereby enabling the
extension of pulsed slow neutron research into new environments.  The
relatively low thermal and radiation loads on the cryogenic neutron
moderator in this type of source enable  new research on cold neutron
moderator materials which may possess higher brightness and/or lower
spectral temperatures than those presently used at existing slow
neutron sources. Operation at a university, made practical by the
minimal activation near the target, also makes LENS  a model facility
for student education and neutron instrument development. The LENS
neutronic design is of broader interest because it possesses several
features similar to those envisioned for future neutron facilities such
as long-pulsed spallation sources (LPSS) and very cold neutron (VCN)
sources. The unique operating regime of LENS also encourages
speculation about the possibility for realization of entirely new types
of neutron sources. We expect LENS to run at a power level of 8 kW or
greater starting in 2008. In the following, we benchmark many of our
simulations to 30 kW of proton beam power, which we expect to be the
practical upper limit for a source of this basic design, limited
primarily by thermal stress in the target.

Although not a spallation source, the LENS source neutronics
nevertheless shares many features with proposed long-pulse spallation
sources since most of the neutrons initially produced in each source
are predominantly in the 1-10 MeV energy range. LENS uses a ``coupled''
moderator (namely, a moderator whose neutron field is directly
correlated in space, time and energy with that in the reflector) to
cool the 1-10 MeV source neutrons to form a cold neutron pulse width on
the order of 1 msec, which is the natural timescale for moderation of
fast neutrons in cold hydrogenous materials to the meV energy regime.
Since this timescale is also matched to the macropulse width from MW
power GeV energy proton linear accelerators, this so-called LPSS
accelerator/moderator combination is an attractive possibility for a
future high-powered spallation neutron source of increased brightness
\cite{mezei1997}. With developments in progress in neutron optics,
neutron chopper technology, and neutron spin echo techniques, a large
fraction of slow neutron scattering spectrometers of interest for
neutron spectroscopy can be designed to accept the $\sim$msec-wide
neutron pulses from such a source \cite{Stride2000,Mezei2000}. Existing
short-pulse spallation neutron sources
\cite{lander1986,Muhrer2004,mason2006,picton2005,harada2005} achieve
narrower pulse widths on the order of 50-100 $\mu$s for $\sim$5 $\AA$
neutrons \cite{carpenter1986} to improve the precision of measurement
of neutron energy by time-of-flight techniques. The time distribution
of neutrons emerging from the moderator, also called the
\textit{emission time distribution}, is then  much shorter than the
time of flight (TOF) from the moderator to the instrument, resulting in
high energy resolution. However this order of magnitude reduction in
the neutron pulse width is obtained at the cost of a corresponding
order of magnitude reduction in the neutron source brightness through
the selective use of neutron absorbers near the cold moderator, called
decouplers and poisons \cite{carpenter1986}, which limit the effective
volume from which thermalized neutrons may enter the beams supplying
neutrons to the instruments.

Significant differences between LENS and a typical spallation source
include the comparative inefficiency of (p,xn) reactions ($\sim10^{-2}$
neutrons/proton compared to $\sim20$ neutrons/proton for GeV spallation
reactions) and a less demanding need for heavy shielding in the forward
direction from the target since the most energetic neutrons at LENS are
1.85 MeV below the $\sim$ 10 MeV proton energy. Due to the short
penetration depth of the $\sim$10 MeV  protons into the Be target (on
the order of 1 mm), the initial source volume for the MeV neutrons is
much smaller than for spallation sources.  The LENS target lies on the
line extended from the instruments through the moderator. This
so-called ``slab'' geometry is more efficient than the ``wing''
geometry used in most spallation
sources\cite{Windsor1981,watanabe2003}, but is more prone to
fast-neutron-induced backgrounds. There are, however, clear
similarities between LENS and the current concepts for a LPSS in the
operating parameters that matter most to neutron instrument design,
namely proton pulse width, moderator coupling, operation frequency, and
neutron emission time distribution.

The LENS neutronic design also possesses some overlap with another
related type of neutron source envisioned as a possibility for the
future: a very cold neutron (VCN) neutron source \cite{VCN2005}.
Although sharing several features with the LPSS concept, the motivating
idea behind a VCN source is to shift the neutron energy spectrum into
the VCN energy regime. The VCN energy regime is loosely defined as a
regime below a few meV whose lower limit is defined in practice by the
point at which the effects of gravity on neutron trajectories are so
great that it is no longer possible to make a recognizable neutron beam
for neutron scattering measurements. Since the statistical accuracy in
(non-interferometric) neutron spectroscopy is directly proportional to
the phase space density of the neutron source and the phase space
density of a neutron field in equilibrium with a source at temperature
$T$ is proportional to $1/T^{2}$ \cite{golub1996}, there is a clear motivation for
lowering the neutron energies if neutron spectrometers are able to make
use of the full phase space density of the beam. Since the neutron
index of refraction increases for lower energy neutrons it may be
possible to design neutron optical elements and spectrometers that
fulfill this condition.

Several details of a possible future VCN neutron source have yet to be
worked out, and it is by no means clear that a source with sufficient
brightness in the VCN range for neutron scattering research can be
built. Nevertheless it is clear on general grounds that such a source
would possess several neutronic design elements similar to those found
in LENS. In common with LENS-type and LPSS-type sources, a VCN source
is envisioned to operate in long-pulsed mode with a coupled moderator.
The low temperature of the moderator needed to produce a VCN energy
spectrum will require cryogenic elements whose neutronic impact on the
target/moderator/reflector (TMR) has also been an important element in
the LENS design. The potential need for a VCN moderator to employ
nuclei with lower neutron absorption cross sections than hydrogen, such
as (partially) deuterated materials, means that the volume of the
moderator may need to be larger than a typical hydrogenous moderator, and the
sensitivity of the VCN moderator brightness to the relative dimensions
of a coupled moderator/reflector arrangement are similar to issues also
investigated in the course of the LENS design. Finally, if research on
cold neutron moderation discovers materials which can efficiently
convert neutrons to the VCN energy range, LENS could operate in the VCN
energy regime and act also as a useful test bed for some aspects of
higher-power VCN sources and instrumentation.

In this article, we summarize the important features of the design of
the TMR at LENS and report measurements of the neutron brightness,
energy spectrum, and emission time distribution obtained while running
the source at $\sim$100 W power. In Section \ref{sec:neutronics} we
outline the selection of materials and other general features of the
LENS design, and in Section \ref{sec:performance} we describe the
computer model used to simulate the source and compare the measured
performance to the predictions of that model. The detailed neutron
transport analyses described in this paper were carried out using the
MCNP series of Monte Carlo transport codes \cite{mcnp5}. We finish with
a description of possible modifications to the system design (Section
\ref{sec:improvements}) that could increase the cold neutron brightness
and summarize with some conclusions (Section \ref{sec:conclusion}).

Historically, many of the relevant concepts for neutron moderation and
transport were developed for nuclear engineering and reactor design in
time-independent configurations and have subsequently been adapted to
describe time-dependent fields encountered in pulsed neutron sources.
In this paper we make use of these concepts and the associated language
but we also strive to explain some jargon from this field which may be
unfamiliar to physicists. We hope that this work might therefore be
useful for a somewhat wider audience of scientists, including those who
may be interested to design and construct their own neutron source. The
interested reader can find treatments of neutron moderation theory in
Glasstone \cite{glasstone1952} and in Williams \cite{williams1966}. A
general overview of neutron sources is found in Carpenter and Yelon
\cite{carpenter1986}, Windsor\cite{Windsor1981} describes many concepts
important for pulsed neutron source design, and an excellent review of
modern spallation source neutronic design may be found in Watanabe
\cite{watanabe2003}.

\section{Neutronics}
\label{sec:neutronics}
\subsection{Overview of the LENS Design Concept}

In brief, the LENS neutron source consists of: 1) a proton LINAC
capable of delivering a pulsed beam of 7 or 13 MeV with a 25
mA peak current and adjustable pulse width and frequency, 2) a
water-cooled Be target, 3) a room-temperature neutron
reflector/moderator, 4) a cryogenic moderator and 5) biological
shielding. This paper concentrates mainly on those aspects of the LENS
source connected with neutron production and moderation. In this
section we present an overview of the source properties and some simple
dimensional considerations before proceeding to the more detailed
neutronic analysis.

The LENS accelerator is capable of producing beam pulse widths from a
few microseconds to more than a millisecond. Details of the proton
accelerator energy, current, pulse structure, and proposed upgrade path
may be found in \cite{Klyachko2006,derenchuk2005,derenchuk2006}.
Initially the facility is using an  existing 7 MeV  Radio Frequency
Quadrupole (RFQ)/Drift Tube Linac (DTL) accelerator \cite{friesel1998}.
Soon a second DTL section will be added to increase the proton energy
to 13 MeV, which will increase the neutron yield per proton from the
target by a factor of about 4. The peak current and cooling limitations
of this accelerator will limit the operation of LENS to 8 kW
initially\cite{derenchuk2006}. The \nuc{9}{Be}(p,xn) reaction's
relatively soft neutron energy spectrum and low production of
gamma-rays per neutron make it feasible to operate LENS at a university
with an acceptable level of safety and security, and without extensive
remote handling facilities for dealing with activated components. The
neutrons produced from (p,xn) reactions in Be are less energetic than
at spallation sources. The energies of neutrons produced in these
reactions are also below the threshold of many  reactions which could
cause activation in the target.  Since the threshold for the
\nuc{9}{Be}(p,t)\nuc{9}Be reaction is 13.4 MeV, and we wish to minimize
target activation, we have chosen a maximum proton energy of 13 MeV.
Gamma production in the target is dominated by the
\nuc{9}Be(p,$\alpha$)\nuc{6}{Li} channel, which produces one 3.56 MeV
gamma for every ten neutrons liberated from the target
\cite{guzek1998}. Fission and spallation sources typically generate
5-10 gamma rays per neutron
\cite{peele1971,pleasonton1972,verbinski1973} directly in the
source/target. Neutron capture in the reflector, and not proton
reactions in the target, is the dominant source of gammas within the
LENS TMR. However, the neutron capture gamma rays are also a more
distributed source of radiation than the gamma rays produced in the
target. The primary construction materials near the target and
moderator are Be and Al, so the long-term gamma field from activated
products at the center of the source is dominated by alloying elements
in these materials. Thus, we expect the activation of TMR materials
have negligible impact on the instrument backgrounds and we expect the
radiological hazard even of these core TMR materials to be low enough
so as to be manageable without the need remote handling.

It is not difficult to estimate some of the relevant length scales and
time scales for the LENS TMR. LENS employs a light-water reflector
coupled to a cryogenic methane cold source in a geometry which
minimizes neutron absorption without unduly broadening the neutron
pulse width. The primary neutrons rapidly slow down to thermal energies
in the water reflector through elastic collisions with the protons in
the water, and then diffuse throughout the TMR until they leak out or
are absorbed. Relevant length scales include (a) the neutron slowing
down length, which is the RMS distance from from the point where a fast
neutron enters the medium to where its energy is lowered to a fraction
of an eV, (b) the diffusion length, which is the distance such a
thermalized neutron diffuses before being absorbed, and (c) the
migration length, which is the quadrature sum of the neutron slowing
down length and the diffusion length. For water, this migration length
is on the order of 6 cm, and therefore we expect to need a reflector
volume with a radius on the order of 20 cm, or three migration lengths.
The cold moderator is fed by the neutrons that thermalize in the
reflector, so the emission time distribution of the moderator is
strongly correlated to the neutron transport properties of the
reflector. Also, since our reflector is large compared to the migration
length, the characteristic decay rate for the neutrons is set by the
absorption rate of the thermal neutrons in the reflector and is given
roughly by \cite{williams1966},
\begin{equation}
\xi = v_{th} \Sigma_a \mbox{ [sec$^{-1}$]} \label{xi}
\end{equation}
where $v_{th}$ is the thermal neutron velocity and $\Sigma_a$ is the
macroscopic absorption cross-section. Given equation \ref{xi}, we see
that the expected decay time for thermal neutrons in a large water
reflector should be about 270 $\mu$sec. Table
\ref{tbl:trans-params1} summarizes the various length scales for a
number of candidate materials considered for the LENS reflector.

\begin{table}
\centering \caption{Neutron transport parameters for candidate
reflector materials \cite{glasstone1952}. \textsl{Absorption length}:
$v_{th}$ times the absorption time; \textsl{Diffusion length}:
diffusion distance for a thermal neutron in an absorption time;
\textsl{Slowing down length}: RMS distance from fast-neutron entry
point to point at which energy falls below 0.4eV, \textsl{Migration
length}: thermal diffusion and slowing-down lengths added in
quadrature; \textsl{Slowing Time}: the average time for a 2 MeV neutron
to slow to a 25 meV thermal neutron.} \label{tbl:trans-params1}
{\scriptsize
\begin{tabular}{p{0.75in}|p{0.5in}p{0.5in}p{0.5in}p{0.5in}p{0.5in}p{0.5in}p{0.5in}}
\hline\hline Material & Density (g/cc) & Absorption Length (cm) &
Diffusion Length (cm) & Slowing Down Length (cm) &
Migration Length (cm) &  Slowing Time ($\mu s$)  \\\hline  %
Light Water & 1.0 & 58 & 2.9 & 5.7 & 6.4& 10  \\\hline %
Heavy Water & 1.1 & $1.25 \times 10^4 $ & 100 & 11 & 103 & 67   \\\hline %
Beryllium & 1.8 & 770 & 23.6  & 9.9 & 25  & 46 \\\hline %
Graphite & 1.6 & 2780  & 50.2 & 19 & 54 & 150
\\\hline\hline
\end{tabular}
}
\end{table}

The moderator is cooled to maximize the flux of long wavelength
neutrons \cite{kiyanagi1995a,oi2004,Inoue1982}.  The lower intensity of
the fast neutron and gamma fields near the LENS target should allow the
LENS moderator to operate at low temperatures without  major
disruptions from the recombination reactions in the methane
\cite{carpenter1987,Kulagin2004} that have caused problems at
spallation sources with operation  $<$20 K. Annealing periods for the
cold source will need to be determined empirically once the source
begins operation at powers in the 10 kW range, but these are expected
to be manageable (see also Section 2.3.3 below).

Biological shielding near the source consists of a matrix of lead,
borax, polyethylene, and epoxy \cite{LWTS2002}.  A material with high Z
is required to attenuate the gamma field produced both in the reflector
and by capture gammas from the inner layers of neutron shielding
\cite{shultis2000}.  The first layer of gamma shielding is a high
purity lead layer separated from the TMR by a thin flexible boron
loaded rubber to reduce the activation of the lead layer
\cite{boroflex}. MCNP predicts subsequent layers of alternating borated
poly and lead will limit contact dose from neutrons and gammas (at 30
kW power) at the TMR surface to 1 REM/h. The TMR is enclosed within a
concrete vault with 1.3 m thick walls, and outside this vault we expect
a biological dose rate of less than 1 mREM/h \cite{ncrp1965}.

\subsection{Neutron Production Target}
The  neutrons are produced by bombarding a  Be target with protons. A
metallic Be target was chosen because of its high neutron yield, high
melting point, and mechanical strength. The range of a 13 MeV proton in
Be is 1.3 mm and we have chosen a 3 mm thick Be target. The short range
of $\sim$ MeV protons in materials also implies the target must be
exposed directly to the proton beam vacuum.

Operation of the target at several kW power will induce thermal
stresses in the target through mechanical deformation during the proton
pulse, time average heating of the target, and the high power density
of the incident proton beam.   The proton beam will be spread
relatively uniformly over an area of $\sim$50 cm$^2$ on the target by 2
octupole magnets to reduce the peak power density applied to the
target. For higher power operations, the target will be bonded on the
back side to an Al substrate cooled with flowing water. The Al plate is
designed to employ the concept of hypervapotron cooling \cite{chen2006}
to dissipate an average thermal load of up to 30 kW (6 MW/$m^2$).
Hypervaportron cooling achieves high cooling power and avoids possible
localized heating of the target from a static vapor layer through
proper choice of coolant flow rate, pressure, and channel geometry to
set up local circulating flows that sweep bubbles away from the
surface. This technique requires only a relatively thin layer of water
behind the target. As a result, we have considerable freedom to choose
the thickness of water between the target and the moderator to optimize
cold neutron production. From the neutronics point of view, this water
can act as a premoderator and/or limit the fast neutrons seen by the
instruments.

Total neutron yields and energy spectra have been collected from the
literature \cite{lone1977,lone1982,brede1989,howard2001} for proton
energies up to 23 MeV.  An empirical formula for the total neutron
yield, $Y_N$, as a function of the proton energy, $E_p$ in MeV, is
\cite{nann2006}
\begin{equation}
    Y_N(E_p)=3.42\times10^{8}(E_p-1.87)^{2.05} \mbox{     [n/$\mu$C]}.
    \label{eq:target_yield}
\end{equation}

This gives $1.6\times 10^{-3}$ n/p at 7 MeV proton energy and
$7.6\times 10^{-3}$n/p at 13 MeV. Normalized energy spectra and angular
distributions of the primary neutrons are required as input for
modeling with MCNP. Since to our knowledge no experimental data on
neutron energy distributions from protons on thick Be targets are
available between 5 and 14.8 MeV proton energy, neutron energy
distributions were calculated for the incident proton energies of 7 MeV
and 13 MeV.

\begin{figure}[!h]
    \begin{center}
    \includegraphics[]{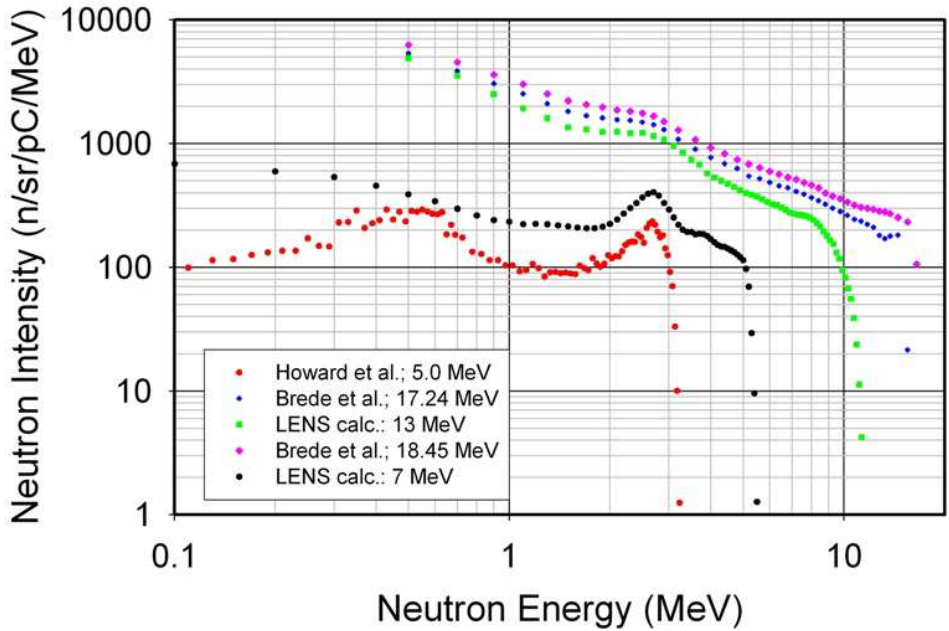}
    \end{center}
    \caption{A comparison of measured \cite{howard2001,brede1989} and calculated \cite{nann2006} neutron
     energy spectra at 0 degrees.}
    \label{fig:srcterm}
\end{figure}

Figure \ref{fig:srcterm} shows measured ``thick'' target neutron
spectral distributions of the \nuc{9}Be(p,xn) reaction at 0$^{\circ}$
for the proton energies of 5 MeV \cite{howard2001}, 17.2 MeV and 18.4
MeV \cite{brede1989} together with the results of our estimates for 7
and 13 MeV, which will be discussed below. As can be seen, a low energy
continuum dominates the spectrum, especially for the higher bombarding
energies. The sharp edge at the high-energy end of the neutron spectra
corresponds to the kinematic limit of the \nuc{9}Be(p,n$_o$)\nuc{9}B
ground state transition.

Calculations of neutron spectral distributions have to take into
account the various reaction mechanisms which contribute to the neutron
production. The \nuc{9}Be(p,n) reaction proceeds through the formation
of the compound nucleus, \nuc{10}B, through direct charge exchange, and
through multi-body breakup (\nuc{9}Be is a loosely bound (2$\alpha$+ n)
system which breaks up very easily). Each process contributes to
different portions of the neutron spectral distributions.
Pre-equilibrium decay of the compound nucleus contributes a high
energy tail to the neutron spectra. The multi-body breakup and
\nuc{9}Be(p,p$^\prime$n) reactions will produce lower energy neutrons
with relatively flat angular distributions. The direct charge exchange
process generally leads to the production of high energy neutrons with
forward peaked angular distributions. However, its contribution to the
total yield decreases steadily with increasing proton energy
\cite{Byrd1983}.

Taking into account the general behavior of the various reaction
mechanisms which lead to neutron emission, we calculated neutron
spectral distributions for a proton bombarding energy of 7 MeV.
Excitation functions of angular distributions of differential cross
sections of the \nuc{9}Be(p,n)\nuc{9}B reaction to the ground and first
three excited states were calculated with the code DROSG-2000
\cite{Drosg1987} in 0.1 MeV steps from 7 MeV down to the threshold for
$\theta $= 0  to 180$^{\circ}$  in 20$^{\circ}$  steps based on data of
Young et al. \cite{Young1990}. For each bombarding energy, these
differential yields were integrated over solid angle and neutron energy
and then subtracted from the total \nuc{9}Be(p,n) cross section
collected by Byrd et al. \cite{Byrd1983} to give the contributions from
the other reaction mechanisms. These other reaction mechanisms of
pre-equilibrium compound nucleus decay and multi-body breakup produce
nearly isotropic angular distributions. They are described as
evaporation spectra, i.e. $ \phi(E) \sim E e^{-E/\tau}$ , where E is
the neutron energy and $\tau$ is an appropriate nuclear temperature.
For each angle, these two distributions were combined with equal
weight, normalized to the above described difference, and then added to
the DROSG-2000 results. After integration over neutron energy and solid
angle this sum was normalized to the total neutron yield given in
Equation \ref{eq:target_yield}.

The neutron spectral distributions for the incident proton energy of 13
MeV were obtained for $\theta $= 0 to 180$^{\circ}$ in $20^{\circ}$
steps by extrapolating the results of Lone et al. \cite{lone1977},
measured at bombarding energies of 14.8, 18, and 23 MeV, and Brede et
al. \cite{brede1989}, measured at 12 energies between 17.2 and 22.0
MeV. This method was considered more accurate than using the above
described method for 7 MeV and extrapolating the data of Young et al.
\cite{Young1990} to 13 MeV. The normalization to the total neutron
yield was done the same way as for 7 MeV.

\subsection{Design of Target/Moderator/Reflector (TMR) System}

We may define three more-or-less separate neutronic functions for a TMR
system: (P) initial production of fast neutrons, (S) slowing down of
neutrons from the $\sim$MeV scale to the $\sim$eV scale through elastic
collisions with light nuclei and delivery to the cold moderator, and
(M) moderation of the thermalized neutrons to lower neutron energies
through interaction with the inelastic modes of the cold moderator. The
information needed for the production (P) step usually either exists
with sufficient precision for neutronic purposes or can be interpolated
or calculated (as we have been forced to do for (p,n) reactions in Be
as described above). Likewise the information needed for step (S) also
exists: cross sections are known, and the energy loss per collision in
this energy regime can be calculated from elementary classical
mechanics. The information needed for step (M) does not always exist
since the low energy inelastic mode structures of condensed matter
systems depend a great deal on the details of the material's structure,
and adequate models are difficult to construct from first principles
\cite{Granada2006a}.

It is interesting to note that for various reasons the separation of
these three neutronic functions is somewhat more sharply defined for a
neutron source of the LENS type than for reactors or spallation
sources. Reactor cold neutron sources possess design constraints
associated with the need to maintain the fission chain reaction in the
reactor, thereby inducing a design coupling between functions (P) and
(M). High-power spallation neutron sources are generally designed with
several moderators close to the target which compete with the space
available for the reflector, thereby effectively introducing a coupling
between functions (P) and (S). Both reactors and spallation sources
produce the initial fast neutrons in an extended volume, thereby
inevitably spatially mixing all three functions, and the intense high
energy neutron and gamma fields introduce another coupling of elements
(P) and (M) indirectly through the cryogenic demands on the refrigerator for the
cold moderator.  For a (p,n) reaction source such as LENS, however,
these functions can be more cleanly isolated. Due to the $\sim$1 mm
range of 10 MeV neutrons in Be, the fast neutrons are emitted from a
sharply-defined narrow plane with a transverse area defined by the
incident proton beam, thereby spatially isolating function (P). Since
neutron beams are extracted from a relatively small solid angle, the
volume of the reflector can be much larger than the volume of the cold
moderator, and to the extent that the cold moderator can be viewed as a
relatively small perturbation to the neutron field developed by the
reflector, a separation of functions (S) and (M) is effected. Finally
the lower fast neutron and gamma intensity of the LENS source reduces
the heat load on the cryogenic moderator to further weaken the coupling
of functions (P) and (M) through cryogenic engineering constraints. The
combination of these features simplifies the design process and allows
us to separately optimize each of the three components as a starting
point for the later refined calculations using realistic geometries.

Since the three main functions noted above possess some independence,
we are freer to specify exactly what we want the reflector and the cold
moderator to do. In this simplified view, the main purpose of the
``reflector'' (in reality a combined reflector and moderator) is to
moderate the fast neutrons into the thermal energy regime and return
the largest possible fraction of them to the cold moderator, and the
main responsibility of the cold moderator is to convert these
thermalized neutrons as efficiently as possible into the cold neutron
regime.  Of course, we also note that all this must be done while
maintaining a time structure from the combined system such that any
long tails present must be acceptably small for the neutron scattering
instruments.

\begin{figure}[!h]
    \begin{center}
    \includegraphics[]{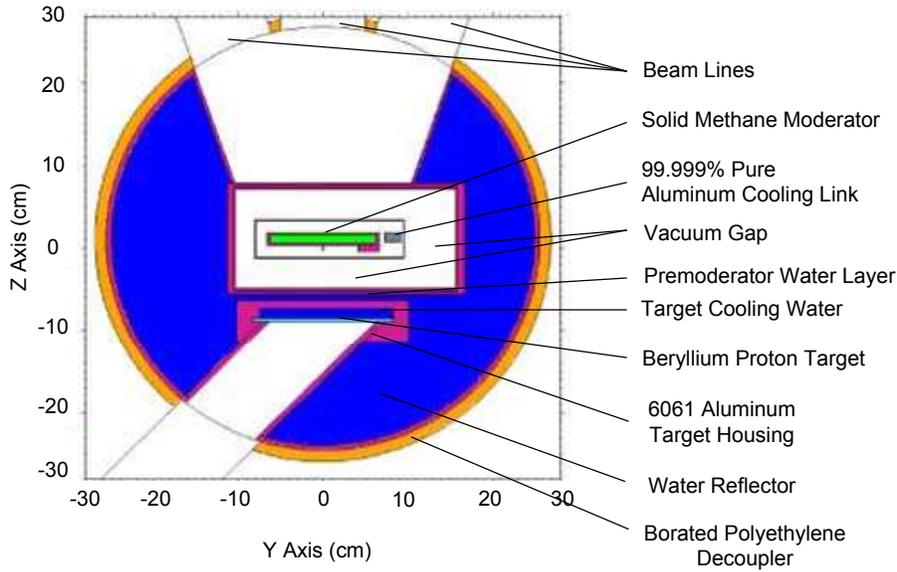}
    \end{center}
    \caption{YZ Planar view of the TMR geometry.  The 1 cm thick methane moderator is at the center,
    forward of the target and target cooling.  The target and moderator are surrounded by a light
    water reflector of radius 25 cm, decoupled from the shielding layers by a 5 cm thick borated
    poly layer.  Biological shielding layers are not shown for
    clarity, but are included in the simulations.}
    \label{fig:tmr2}
\end{figure}

\begin{figure}[!h]
    \begin{center}
    \includegraphics[]{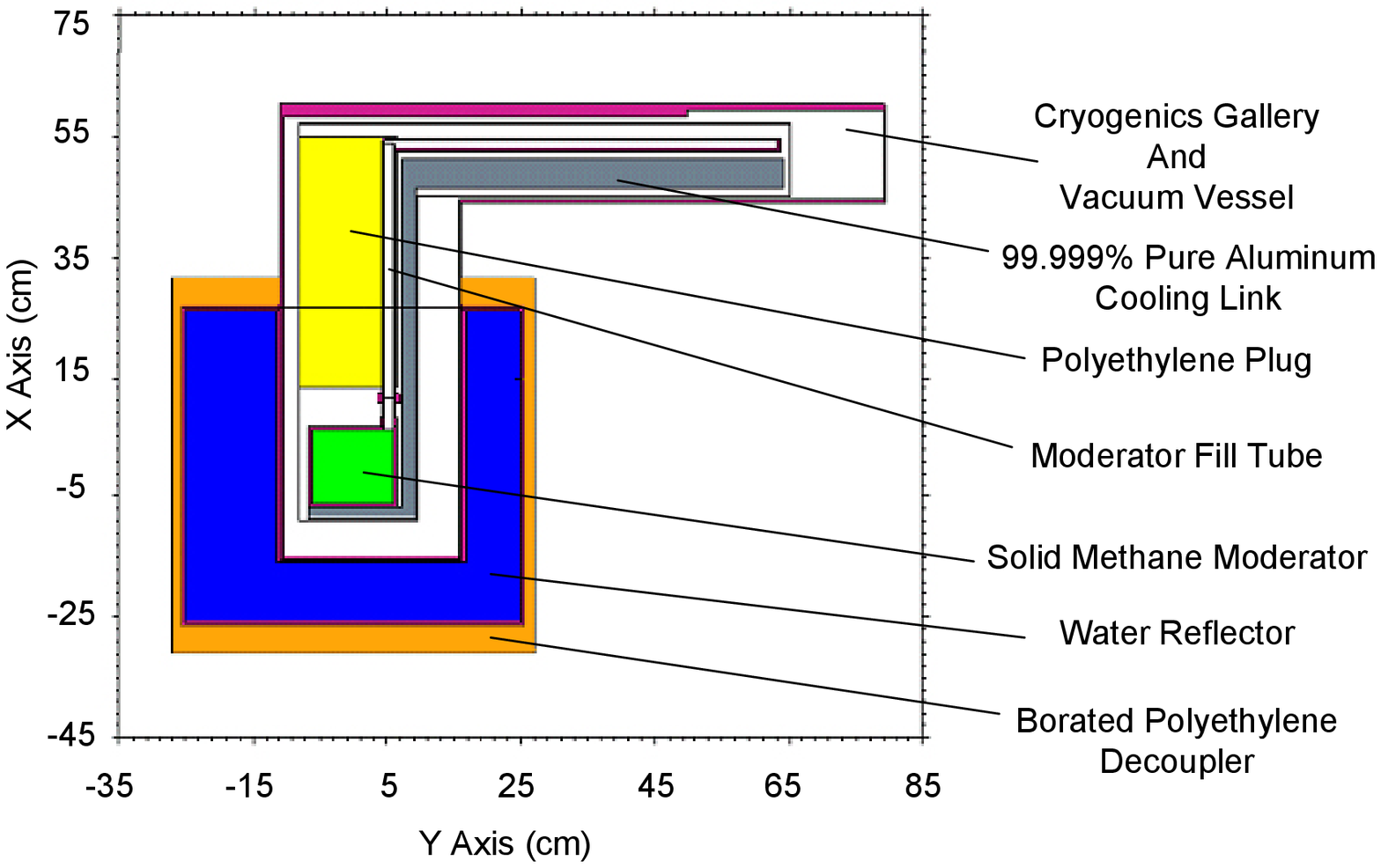}
    \end{center}
    \caption{YX Planar view of the TMR geometry.  The 12x12 cm methane slab is
    at the center, connected to the helium refrigerator (not shown) by a 99.999\%
    pure aluminum cold link.  The water reflector surrounds the moderator
    and the polyethylene plug above the moderator provides neutron reflector
    material inside the cryogenic system. Biological shielding layers are not shown for
    clarity, but are included in the simulations.}
    \label{fig:tmr1}
\end{figure}

The MCNP model geometry of the TMR is shown in Figures \ref{fig:tmr1}
and \ref{fig:tmr2}.  The primary neutronic elements of the source are a
cylindrical light water reflector surrounding the beryllium target and
a relatively thin solid methane moderator, 12x12x1 cm$^3$. The mean
free path (MFP) of a neutron with energy between a few eV and several
keV in solid methane is on the order of 1 cm, and drops to 0.1 cm for
thermal neutrons \cite{macfarlane1997,grieger1998}.  Thus, the methane
moderator interacts relatively weakly with the fast neutron field and
more strongly with the slow neutron field as required to justify our
consideration of reflector and moderator functions separately.

The present TMR design includes extra space around the moderator in
order to facilitate experimental studies of various neutronic changes
(such as changes in premoderator, moderator geometry, poisons, etc.) at
the present very low power levels. We show later in this paper that a
gain of a factor of up to 1.3 in neutron moderator brightness at a
given proton beam power is possible if this vacuum space is reduced
considerably.


\subsubsection{Moderator geometry}
MCNP calculations using a simplified model of infinite plane slabs of
finite thickness 22 K solid methane moderator were performed to
determine the dependence of cold neutron brightness on methane
thickness using the \emph{smeth22k} kernel \cite{macfarlane1997}.  If
cold neutrons are to be produced directly from the primary fast neutron
flux with negligible premoderation, as is true with most spallation
source designs to date, then the optimal thickness of methane is about
5 cm, as one finds in existing and planned methane moderators at
spallation sources such as IPNS and ISIS.

\begin{figure}
  \begin{center}
  \includegraphics[]{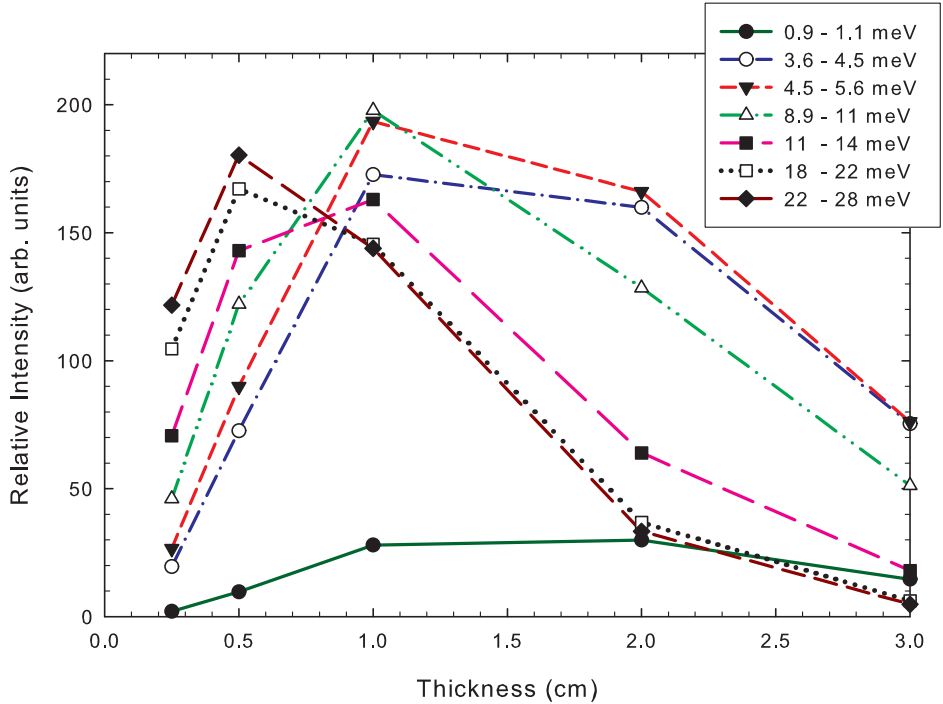}
  \end{center}
  \caption{Simulated intensity of cold neutrons emitted from the exit face of
  an infinite slab, finite thickness methane moderator for a 100 meV pencil neutron beam incident on one side.
  A broad 1-2 cm maximum is observed for production of cold neutrons .}
  \label{fig:slow_neutron}
\end{figure}

However, we found that if the moderator is designed to couple primarily
to the thermal neutron field  from the reflector, as in our geometry,
then the methane can be  thinner, just 1 or 2 cm. This has important
consequences for moderator cryogenics: a thinner moderator greatly
reduces the heating from fast neutron energy loss in the moderator
medium and also reduces the temperature gradient across the thickness
of the moderator medium, which possesses poor thermal conductivity at
low temperature. Figure \ref{fig:slow_neutron} shows the relative
probability to produce cold neutrons in different energy ranges from a
monoenergetic beam of incident neutrons of 100 meV. A broad maximum in
cold neutron intensity is observed from 1 to 2 cm methane moderator
thickness. To minimize neutron and gamma heating and the thermal
gradients, 1 cm was chosen as the thickness of the methane moderator.

More detailed simulations of the emitted cold flux in the full
neutronic model using the \textit{smeth22K} kernel confirmed that
little cold flux would be gained by going beyond 1 cm thickness and
also showed that cold neutrons are emitted from the moderator surface
with a flat spatial distribution for our design.  More recent
simulations with an improved methane scattering kernel suggest that a
thickness of 2 cm may produce a slightly greater cold neutron
brightness for methane temperatures less than 20 K \cite{shin2005}.
Both MCNP studies confirm the initial insight that the LENS methane
moderator can be significantly thinner than found at existing
spallation facilities.

Given the size of the neutron beam lines, no more than $10\times10$
cm$^2$, it was found that cold neutron flux leaving the moderator's
face (the ``leakage flux'') saturates near 12$\times$12 cm$^2$
moderator area and so this size was taken for the lateral dimension of
the slab moderator.  In order for a neutron to be seen by the
instrument, the neutrons incident on the cold moderator must diffuse
into a region of the moderator surface viewed by the beam line.  The
additional width can be understood as the length a low energy neutron
from the reflector can travel as it thermalizes in the methane and
still contribute to the cold neutron flux.  As can be seen in Figure
\ref{fig:slow_neutron}, the contribution from thermal neutron energies
($>$10 meV) falls off rapidly for distances greater than 1.0 cm.

\subsubsection{Reflector material}
\begin{figure}
  \begin{center}
  \includegraphics[]{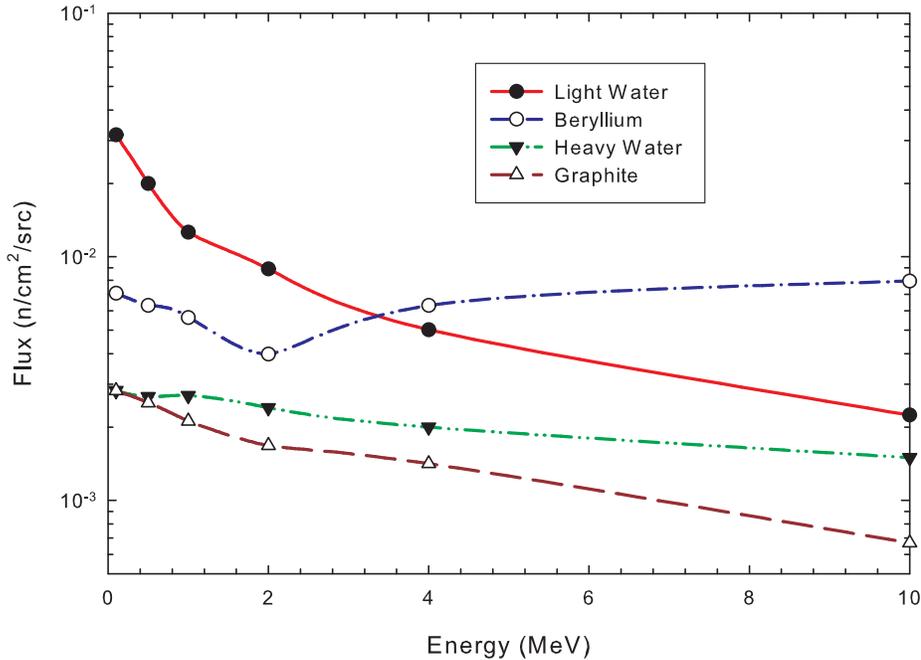}\\
  \end{center}
  \caption{Simulated intensity of thermalized neutron flux per source neutron in the center of spherical ambient temperature
  reflector materials, radius 40 cm,  as a function of the energy of a monoenergetic point source also at the center.}
  \label{fig:reflector}
\end{figure}

For a neutron source with strong coupling of the thermal neutron field
between the moderator and the reflector, the reflector is actually
being employed as both a reflector \textit{and} a thermal neutron
source \cite{kiyanagi1995a,kiyanagi1992,kiyanagi1995b}. Therefore we
first considered calculations of reflector materials in model spherical
geometries to investigate the ability of various materials to produce
high thermal neutron flux in the center of the reflector as a function
of primary neutron energy and reflector radius.  While many different
radii were simulated, a representative result that conveys the general
thrust of the calculation is shown in Figure \ref{fig:reflector}.  In
this figure, 40 cm radius spheres are used with an isotropic point
source of fast neutrons at the center. As we have discussed, our design
attempts to thermalize the fast neutrons rapidly in the reflector and
allow thermal neutrons to diffuse back to the moderator for efficient
cold neutron production. For primary neutron energies lower than about
3.5 MeV, light water is the best reflector material. As the primary
neutron energy increases above 3.5 MeV, the neutron-hydrogen scattering
cross-section decreases, thereby increasing the  mean free path and
reducing the fraction of thermalized neutrons returned to the
moderator. In contrast, for neutron energies above the 3.5 MeV mark
(about 25\% of the 13 MeV proton source primary yield), Be is a better
reflector for a coupled cold neutron source.  This is due to its larger
neutron scattering cross section and also the extra neutron production
in Be from (n,2n) reactions (which multiplies the primary flux by a
factor of 1.09). Although these calculations indicated that Be exhibits
better performance, light water was chosen for the initial LENS
reflector to minimize both cost and potential activation issues.

We selected a cylindrical reflector  of light water 25 cm in radius, 50
cm tall enclosing the production target and moderator based on these
simulations. Although no cold flux gain is realized from increasing a
water reflector radius beyond 20 cm, calculations indicate that the
optimal beryllium reflector radius is 25 cm and we did not want to
preclude the possibility of an upgrade to a Be reflector in the future
(see Section \ref{sec:improvements}).

\subsubsection{Moderator and cryogenics radiation load}

The refrigerator that maintains the moderator temperature must remove
the heat deposited by fast neutrons and gammas. The gamma rays come
from both the production target and from neutron capture on the
moderator, reflector, shielding, etc. (see Table
\ref{tbl:heatingbreakdown}). We chose a CryoMech PT410 commercial pulse
tube mechanical refrigerator \cite{cryomech} to provide the moderator
cooling at LENS.  The thermal connection between the moderator vessel
and the refrigerator consists of a 99.999\% pure Al plate connected to
the bottom of the moderator vessel and coupled to the 2$^{nd}$ stage of
the refrigerator \cite{Al5N}. The total length of the Al connection is
on the order of 1 m in order to take the cold-head far enough away from
the source to avoid significant activation of its components. Bench
tests with an electrical heater connected to the moderator vessel
showed that temperatures below 10 K may be maintained provided that the
heat load on the moderator is less than 3 Watts.

\begin{table}[!h]
\label{tbl:heatingbreakdown} \centering \caption{Neutron and Gamma heat
loads on the cryogenic components for 30 kW proton beam power at 13 MeV
as calculated in MCNP.  There is an additional 180 mW of heating from
decay activity in the aluminum components which was calculated
separately. An asterisk (*) denotes components that are thermally
connected to the (warmer) first stage of the two-stage mechanical
refrigerator. All other components are connected to the (colder) second
stage.}\label{tbl:heatingbreakdown}{\scriptsize
\begin{tabular}{|p{1.25in}|p{1.0in}|p{1.0in}|p{1.0in}|p{0.4in}|}
  \hline\hline
  \textbf{Cryogenic Element} & \textbf{Neutron (mW)} & \textbf{Capture Photon (mW)} & \textbf{Target Photon
  (mW)} & \textbf{Total (mW)}
  \\\hline
  Solid Methane & 420 & 23 & 36 & \textbf{479} \\\hline
  Moderator Vessel & 26 & 83 & 127 & \textbf{236} \\\hline
  Fill Tube (lower) & 8 & 20 & 27 & \textbf{55} \\\hline
  Fill Tube (upper) & 1 & 6 & 2 & \textbf{9} \\\hline
  Aluminum Bar & 12 & 52 & 44 & \textbf{108} \\\hline
  Moderator Attachment Flange & 1 & 4 & 3 & \textbf{8} \\\hline
  Poly Plug* & 267 & 122 & 43 & \textbf{432} \\\hline
  Thermal Shield* & 16 & 72 & 75 & \textbf{163} \\\hline
  \textbf{Total (mW)} & \textbf{750} & \textbf{382} & \textbf{357} &
  \textbf{\textit{1490}}\\
  \hline
  \hline
\end{tabular}
}
\end{table}

The thermal load on the various components of the cryogenic system was
calculated in MCNP by tallying energy deposition from neutrons and the
gamma rays produced both directly in the target and by neutron capture.
The contribution to the whole thermal budget from each component is
shown in Table \ref{tbl:heatingbreakdown}. A cross-sectional area of
$1.8 \times 2.0$ cm$^2$ for the aluminum cooling link produced a total
deposited power of 1.5 W into the cryogenics, of which only about 1 W
is thermally connected to the second (colder) stage of the
refrigerator. Decay gamma and beta heating from activated aluminum is
not included in the MCNP calculations, but was estimated from the
calculated volume averaged neutron flux in each element to be
approximately 180 mW additional power over the course of long term
running at full accelerator power. This heat contribution is dominated
by $^{28}$Al decay in the moderator vessel and thermal link.  Since the
combined load from these sources is well below the 3 W used in the
bench test mentioned above, we are confident that even at 30 kW
operation we will be able to keep the moderator at well below 10 K with
the existing cryogenic design.

These simulations estimate the specific thermal load on the methane
itself to be only about 3.3 mW/cm$^3$. At temperatures between 4 and 10
K the thermal conductivity of solid methane is on the order of 10 $\pm$
5 mW/cm.K \cite{Jezowski1997}, so on dimensional grounds we can expect
to see gradients on the order of only 0.3-1.0 K/cm through the
thickness of the methane even at 30 kW proton power. Therefore the 1 cm
thick moderator at LENS will not develop a substantial temperature
gradient under foreseen operating conditions. The need for aluminum
foam or similar structures used at IPNS to improve thermal contact
between the solid methane and the aluminum walls of the moderator
vessel therefore disappears \cite{carpenter1985b}.

\section{Performance - Simulated and Measured}
\label{sec:performance}
The brightness, flux, emission time distribution, and energy spectrum
of neutrons emitted from the moderator are obviously essential
parameters for neutron instrument design. The emission time
distribution from a long-pulse neutron source such as LENS is broad
enough to significantly perturb the normally straight-forward
relationship between time-of-flight and neutron energy and so we
consider this effect first. Only after understanding the effects of the
broad emission time distribution is it possible to perform quantitative
comparisons of the brightness, flux and energy spectrum with
simulations.

\subsection{Emission Time Distributions}
\subsubsection{Simulation}
The emission time distribution is simulated by tallying the neutron
leakage flux over the instrument side of the cold moderator face
convolved with a square proton pulse as employed in the experiments
(150 $\mu$s for the results quoted here).  A fairly broad cone
($\cos\theta>0.95$) was accepted in the calculation to enhance Monte
Carlo statistics while still retaining sufficient fidelity in
representing the conditions of the experimental measurement.  This is
permissible since  the emission time distribution, which for our
strongly coupled long-pulse source is dominated by the time-dependent
neutron transport properties in the reflector rather than the cold
moderator geometry, is a weak function of emission angle across this
angular range.

\begin{figure}
    \begin{center}
    \includegraphics[ ]{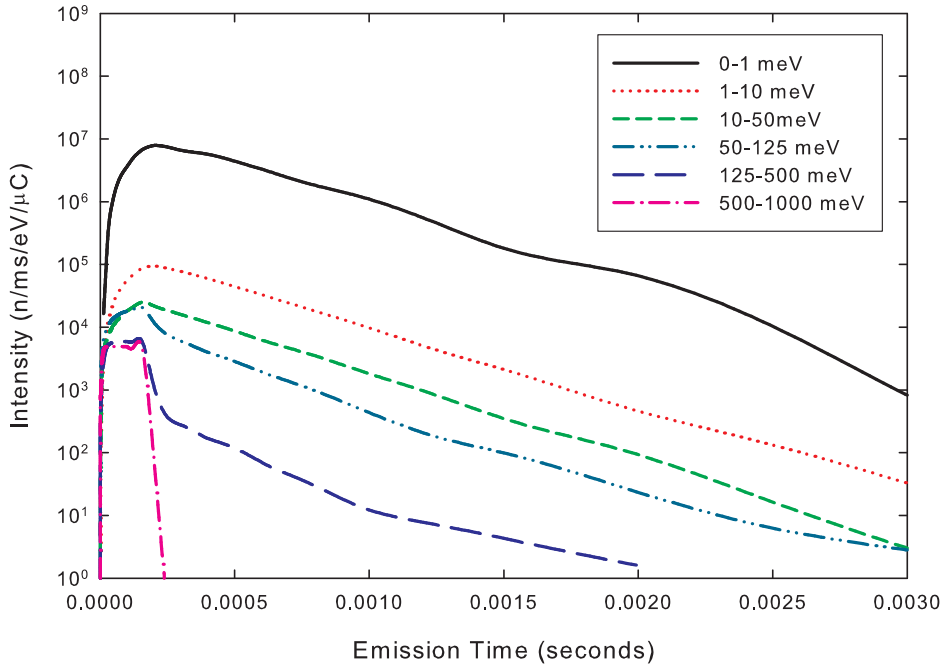}
    \end{center}
    \caption{Simulated neutron emission time distributions.  The water reflector
    and coupled 22 K solid methane moderator create long emission times for all low energy
    neutron groups.  The 150 $\mu$s proton beam pulse dominates the pulse shapes
    of higher energy neutrons (500-1000 meV in this figure).}
    \label{fig:emission_time}
\end{figure}

\begin{figure}
    \begin{center}
    \includegraphics[ ]{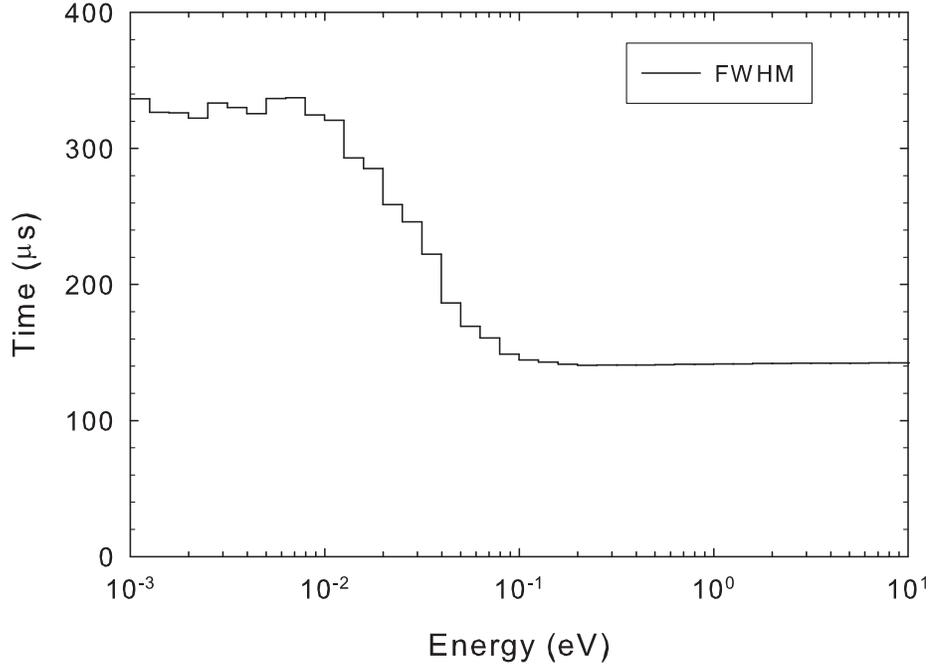}
    \end{center}
    \caption{The simulated average emission time FWHM
    for neutron pulses from a 22 K methane moderator and with a 150 $\mu$s
    proton pulse plotted as a function of neutron energy.}
    \label{fig:offset}
\end{figure}

Figure \ref{fig:emission_time} shows the emission time distribution for
several low energy neutron groups, and the pulse width (FWHM)  is shown
in Figure \ref{fig:offset} as a function of energy.   There is a
saturation in the FWHM near 350 $\mu$s as the neutrons come into
equilibrium with the moderator. The neutron emission rises quickly and
subsequently decays over a time period that is consistent with the
characteristic decay time in the water reflector. We note that this 350
$\mu$sec figure is consistent with the expectations outlined above for
a spectrum dominated by a large water reflector/moderator and a 150
$\mu$sec proton pulse width.


\subsubsection{Measurements}
The emission time distribution was measured using a time-focused
crystal analyzer spectrometer, which is designed to cancel to first
order geometrical effects due to variations in neutron flight path in
the plane of the reflection \cite{carpenter1970,carpenter1985}.  The
crystal monochromator employed a Ge mosaic crystal with mosaic width of
$\sim \frac{1}{2}$ degree, resulting in an energy resolution of $\sim 3
\%$ using the (111) reflection planes. The first order reflection was
set for 2.74 meV (5.46 $\AA$) and the third order reflection was set
for 24.6 meV (1.82 $\AA$);  the second order reflection is forbidden.
These energies were selected because the first order reflection is near
the peak of the cold neutron energy spectrum and the third order
reflection is in the thermal neutron range. These neutron energies span
the extremes of expected emission time distribution FWHM's shown in
Figure \ref{fig:offset}.

\begin{figure}
    \begin{center}
    \includegraphics[ ]{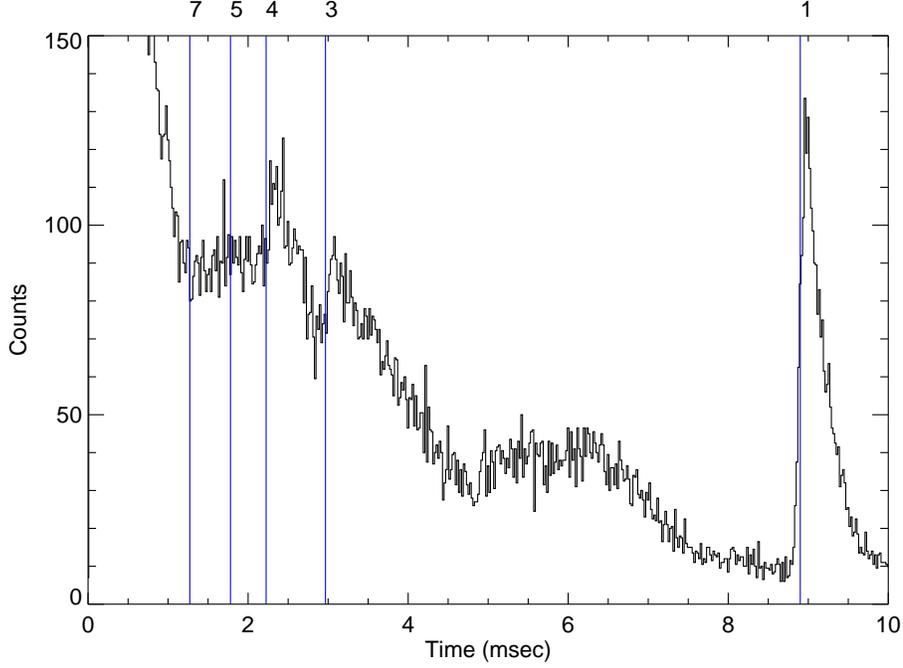}
    \end{center}
    \caption{Raw time-of-flight data for Bragg reflections from the (111) reflection planes in a Ge
    mosaic crystal in a time focussed geometry.  The order of reflection is
    shown at top.  Because the energy resolution of the spectrometer is narrow  compared to the    emission time distribution, the range of energies as reflected in the shape and width of the reflected Bragg peaks gives directly the emission time distribution and FHWM.}
    \label{fig:et_raw_data}
\end{figure}

\begin{figure}
    \begin{center}
    \includegraphics[width=3in ]{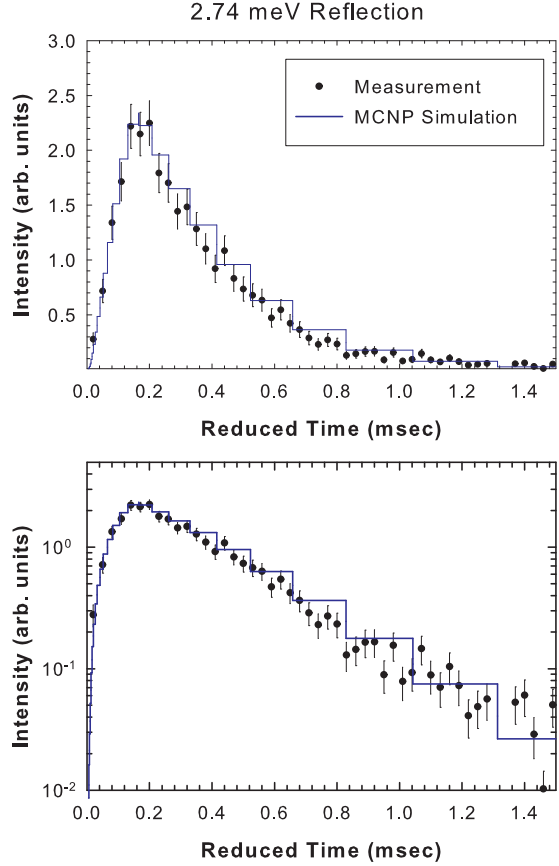}
    \end{center}
    \caption{Emission time distribution  measurement compared with MCNP simulation convolved with 150 $\mu s$
    square pulse.  The time axis is reduced so that each pulse starts at zero time
    and the integral area is normalized to 1 for both curves. }
    \label{fig:et_analyzed}
\end{figure}

The accelerator parameters for this measurement were a 150 $\mu s$
square pulse width, a 7 mA peak current, and 15 Hz pulse rate for a
time-averaged power of 110 W.  The analyzer crystal viewed the
moderator through a beam line oriented at 20 degrees off the moderator
normal. This particular beam line had a 2-cm aperture at a distance of
135 cm from the moderator and the beam was much larger than the crystal
at the analyzer position. Given the broad peaks to be measured, it was
not necessary to fine-tune the detector orientation to optimize the
time-focussing. We estimate that the instrumental broadening of the
measured pulse width is less than 20 $\mu$sec. The total flight path
was 6.6 meters from the moderator to the detector (moderator to
analyzer distance of 6.0 m). The raw data from 12 hours of data
collection is shown in Figure \ref{fig:et_raw_data}. The emission time
distribution at the energy of the first Bragg peak of 2.74 meV agrees
well with MCNP simulation convolved with a 150 $\mu s$ square proton
pulse as shown in Figure \ref{fig:et_analyzed}.

\subsection{Neutron Energy Spectra}
\subsubsection{Simulation}
\begin{figure}
    \begin{center}
    \includegraphics[ ]{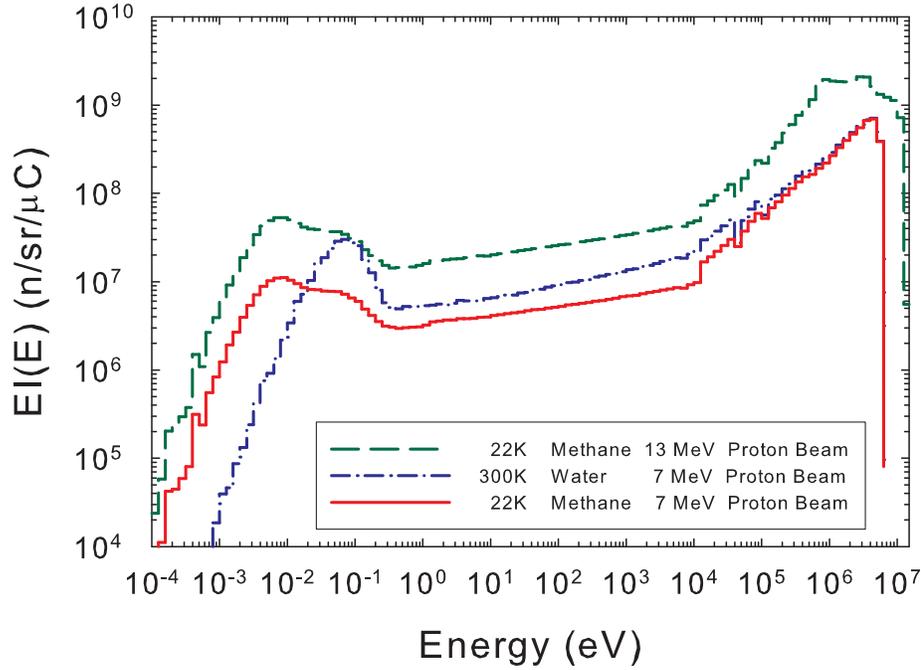}
    \end{center}
    \caption{Neutron spectral intensity predicted by MCNP
    for different proton beam energies and TMR configurations.
    The discontinuity at 10 keV is due to the fact that the finite
    probability distribution used to model the primary neutron flux
    is not defined below this energy. As discussed later, the finite slope
    at epithermal energies may be associated with thin regions of the
    reflector around the cold moderator (see also fig. \ref{fig:gap_spectra}.)}
    \label{fig:spectra}
\end{figure}

Simulated neutron energy spectra (Figure \ref{fig:spectra}) were
calculated for a beam line viewing the moderator at $20$ degrees from
the moderator surface normal. We simulated the energy spectra for three
different configurations of the TMR and proton energy:  a) 7 MeV proton
energy with an empty moderator vessel, b) 7 MeV proton energy with a 22
K methane moderator, and c) 13 MeV proton energy with 22 K methane
moderator. Neutron moderation from fast to thermal energies, which
occurs through a series of non-relativistic 2-body collisions of a
neutron of mass $m_n$ and energy $E$ with a target nucleus of mass $A$
(in units of neutron mass) and energy zero, produce a 1/E neutron
energy spectrum in an infinite medium \cite{carpenter1986}. It is
therefore useful to define a quantity, EI(E), called the spectral
intensity, that is constant in the epithermal energy range for an
infinite medium. The spectral intensity may be related to a measured
neutron flux, $\phi$(E), through the relation:
\begin{equation}
EI(E)=\frac{L^2}{i}E\phi(E) \label{eqn:intensity}
\end{equation}
where E is the neutron energy, $I(E)$ is the luminous intensity  in
units of $n/sr/\mu C/eV$, $L$ is the flight path length in units of
$cm$, and $i$ is the time averaged proton current.  Integral yields, Y
in units of $n/sr/\mu C$, are determined from equation
\ref{eqn:integral} for different energy groups and tabulated in Table
\ref{tbl:results} where they are compared to experimental results.

\begin{equation}
    Y=\frac{L^2}{i}\int_a^b \phi(E) dE = \int_a^b I(E) dE
    \label{eqn:integral}
\end{equation}

\subsubsection{TOF Measurement}
The standard technique for measuring moderator spectra
\cite{iverson1998, micklich2005} relates the count rate in the detector
to the flux through:
\begin{equation}
    C(t)=\Delta T A\varepsilon(E)\phi(E)\frac{dE}{dt}
    \label{eqn:measconv}
\end{equation}
where $C(t)$ is the count rate in the detector at time $t$ after the
start of the proton bombardment of the target, $A$ is the area of the
neutron beam on the detector, $\Delta T$ is the time channel width, and
$\phi(E)$ is the neutron flux at the detector. We used a thin $^3$He
detector manufactured by LND \cite{LND}. The detector efficiency is
determined by the pressure and thickness of the $^3$He absorber:
\begin{equation}
    \varepsilon(E)=1-e^{-n \sigma \frac{\lambda}{\lambda_o} x }=1-e^{-k\lambda} \\
\end{equation}
where $n$ is the number density of the neutron absorber, $x$ is the
absorber thickness, $\lambda$ is the neutron wavelength wavelength, and
$\sigma$ is the absorption cross-section specified at $\lambda_o$.
Efficiency is linear in wavelength when the detector is ``thin'', i.e.
$k\lambda \ll 1$.
\begin{equation}
    \varepsilon(E)\sim k\lambda
\end{equation}

At LENS the simple relation between neutron energy and $t$ is blurred
by the long neutron emission time of the coupled moderator and the long
width of the proton pulses. To completely account for this effect on
the spectra, it would be necessary to deconvolve the energy-dependent
neutron pulse shape from the measured data. For simple spectral
measurements, the dominant contribution to this correction can be
accounted for by introducing an average delay from the start of the
proton pulse to the first emission of neutron of a given energy from
the moderator face. We therefore follow Ikeda and Carpenter
\cite{carpenter1985} and define an average emission time delay,
$\bar{t}(t)$. The dependence of this quantity on neutron energy is
reflected in a dependence on arrival time at the detector, t. This
emission time delay is required to determine the correct mean neutron
energy in a TOF channel, $<E(t)>$, as shown in equation
\ref{eqn:ave_E}.

\begin{equation}
     <E(t)>=\frac{1}{2}m(\frac{L}{t-\bar{t}(t)})^2 \label{eqn:ave_E}
\end{equation}
\begin{figure}
    \begin{center}
    \includegraphics[ ]{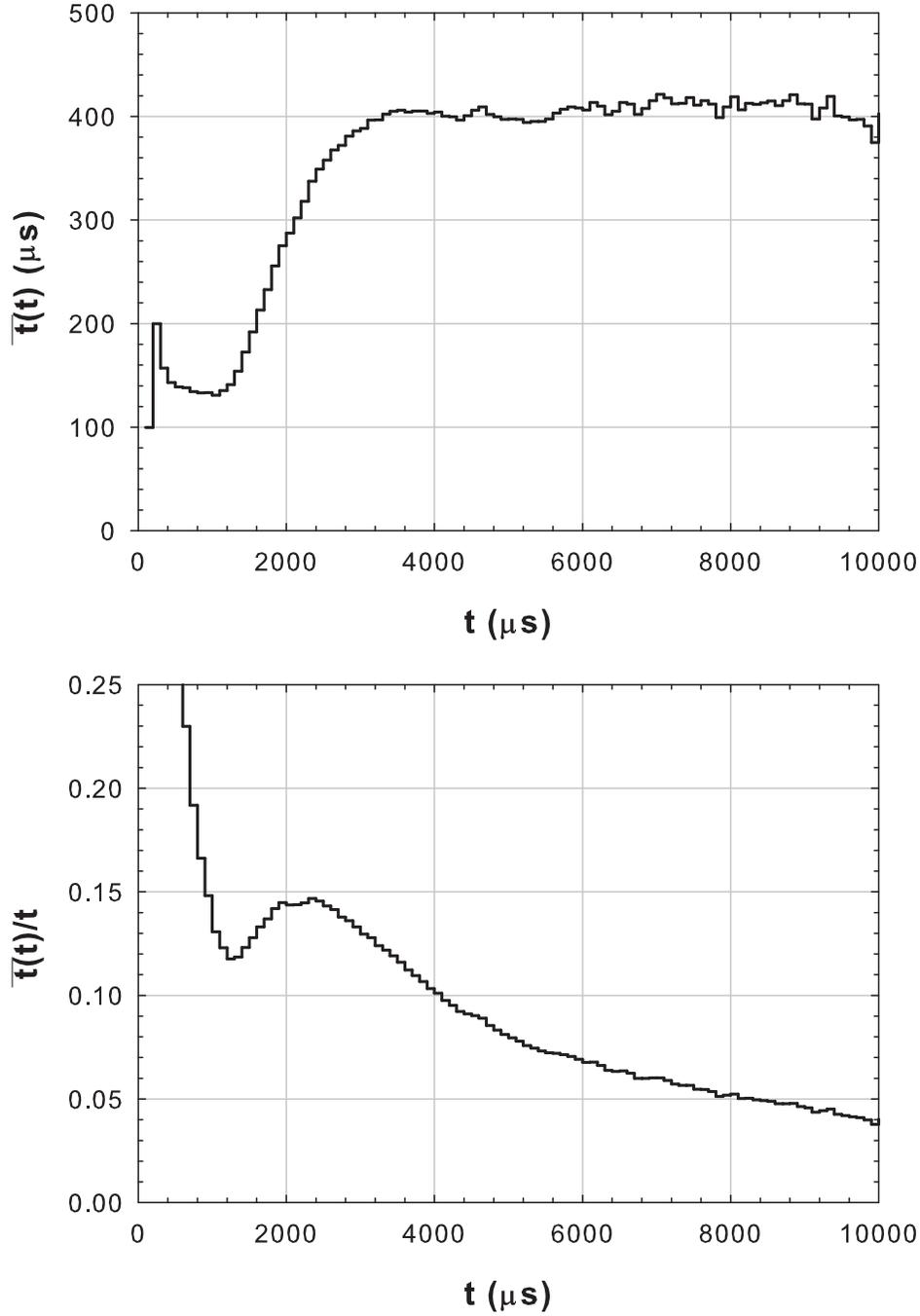}
    \end{center}
    \caption{The simulated average emission time delay, $\bar{t}(t)$ from 22 K methane
    moderator. $t$ is the time elapsed after the start of the
    proton pulse.  The simulations use a 150 $\mu$s proton pulse and
    calculate the flux at a point 570 cm from the moderator, 20 degrees to the moderator normal.  Note the
    delay can be as much as 15\% of the time of flight for energies less than 50 meV.}
    \label{fig:t(t)}
\end{figure}

This correction is typically small at a SPSS where $v\bar{t}$ is a
small fraction of the flight path. At LENS, however, this correction is
large for thermal neutrons: $v\bar{t}$ is $\sim 90$ cm for 200 meV
neutrons, and $\sim 50$ cm for 10 meV neutrons, both of which are
significant fractions of the 570 cm flight path. This is reflected in
Figure \ref{fig:water_flux}, where the uncorrected data deviates from
the corrected data for energies $>$ 30 meV.

We use MCNP simulations to determine the mean energy in a TOF channel
by calculating the integrals in Equation \ref{eqn:calc}.
\begin{equation}
    <E(t_i)>=\frac{\int_{t_{i-1}}^{t_{i}}dt \int_0^\infty dE
    \varepsilon(E) E \phi(E,t)}{\int_{t_{i-1}}^{t_{i}}dt \int_0^\infty dE \varepsilon(E)\phi(E,t)}
    \label{eqn:calc}
\end{equation}

where $t_i$ is the $i^{th}$ time channel upper bound, and $\varepsilon$
is the detector efficiency, which in this case is given by the
cross-section for the $^3$He(n,p) reaction. $<E(t)>$ is calculated at
the position at which spectra are taken in the measurement, and $\Delta
T=t_{i}-t_{i-1}$ is chosen to reflect the 100 $\mu s$ channel width
employed in the experiment. This may then in turn be used in Equation
\ref{eqn:ave_E} to find $\bar{t}(t)$. The final result for $\bar{t}$ is
shown in Figure \ref{fig:t(t)} as a function of $t$.

\subsubsection{Detector Efficiency Calibration}
\label{sec:au-det} The absolute neutron flux was measured using the
technique of gold-foil activation. Both bare and cadmium-covered gold
foils were activated at the detector position simultaneously with the
spectral measurements to provide an absolute calibration of the neutron
detector efficiency \cite{micklich2005,iverson1998}. The reason for
employing cadmium in this type of measurement is easy to understand.
Due to a low-lying resonance in \nuc{113}Cd, the transmission through a
Cd foil is essentially a step function with a cutoff around 550 meV.
Subtracting the Cd-covered activity from the activity of a bare foil
allows one to reduce the impact on the activation per unit volume of
the bare gold foil, $\tilde{A}_{bare}$,  from that higher energy part
of the spectrum which is not easily measured in the TOF experiment.

The saturation activities for the two gold foil measurements may then
be related to the flux through the relations \cite{knoll,astme262}:
\begin{equation}
\tilde{A}=\tilde{A}_{bare}-\tilde{A}_{cd}=\int_0^\infty dE \sigma(E)
\Delta_{Cd}(E) \phi(E) \label{eqn:actbootstrap}
\end{equation}
Where we have for cadmium absorption:
\begin{equation}
\Delta_{Cd}(E)=1-e^{-\Sigma_{cd}(E)x}
\end{equation}

The foil activity was measured with a $\beta-\gamma$ coincidence method
that allows absolute activity to be measured and absolute efficiencies
for the gamma and electron detectors to be determined without a need
for independent standard sources to provide the calibration
\cite{coin}. The foils were placed between a lithium-drifted germanium
detector, used for $\gamma$ detection, and a plastic scintillator
coupled to a photo-multiplier tube used as the $\beta$ detector.

The apparatus was tested using a Au test foil activated closer to the
source, the activity of which was measured periodically during a
$\sim$2 week decay period from an original activity of about 100 Bq to
a final activity below 1 Bq. The estimated uncertainties throughout the
measurement, which increase as the foil decays, were consistent with the
spread of the points around the average, and also consistent when
evaluated against more precise measurements taken on
other foils at even higher activities. At the lower activities, the
uncertainty has significant contributions from both counting statistics
and uncertainty in the background count rates. The measurements were
performed at a facility with an operating 200 MeV cyclotron, and
backgrounds were found to vary by 3 to 5\% depending on the state of
that accelerator. Therefore, as an added precaution, measurements were
taken in five-minute time bins which were histogrammed and examined as
a function of time to check for any anomalous readings or inconsistent
trends in count rate before they were summed. For measurements with the
foils in this study, no such anomalies were seen.  The $\gamma$
background signal in the $\beta$ detector was measured using aluminum
filters and a similar high-activity gold foil. With such an
arrangement, the absolute efficiencies were determined to be about 4\%
in $\gamma$ detection and about 13\% in $\beta$ detection.

At the end of the 10 hour irradiation at the sample position, the bare
foil had activity of 3.7 Bq and the cadmium covered foil had activity
of 1.0 Bq. The uncertainty in the absolute activity is approximately
12\% for the bare foil using the coincidence counting method. This
uncertainty estimate is in good agreement with the differences between
the activities calculated independently using the absolute gamma and
beta detector efficiencies.

These data are used to determine a detector efficiency in the following
manner. The data are converted to the energy domain using Equation
\ref{eqn:measconv} with the $\bar{t}$ correction for the energy. Then
the unnormalized flux is weighted with the activation cross-section and
the product integrated up to the maximum energy that can be measured
with TOF to determine $k$. The proton pulse terminates at $150 \mu s$,
so the highest energy that can be reliably measured via TOF is about 2
eV. To perform the integration over the high energy regime we use a
modified $1 \over{E^{1+\alpha}}$ slowing-down energy dependence to
extrapolate to energies where the measured spectrum is unavailable. The
leakage exponent, $\alpha$, which accounts for deviation from the 1/E
behavior due to losses through the boundaries of a finite moderating
volume \cite{carpenter1986}, is determined from a fit over that part of
the epithermal spectrum that was measured.

The flux used in the numerical integration of Equation
\ref{eqn:actbootstrap} is described by:
\begin{eqnarray}
\phi(E)&=& \phi_{th}(E)=\frac{C(t)}{\Delta T A \lambda \frac{dE}{dt}}
\mbox{\hspace{5mm}$E <E_{max}$} \nonumber \\
&=& \phi_{epi}(E)=\phi_s \frac{1}{E^{1+\alpha}}
\mbox{\hspace{5mm}$E_{max} < E < 5.2 MeV$}\nonumber\\
&=& 0 \mbox{\hspace{5mm}$E>5.2 MeV$}
\end{eqnarray}
Where $\phi_s$ is a constant related to the magnitude of the total
epithermal flux chosen such that the flux is a continuous function of
energy, $\Delta T$ is the TOF bin width, and A is the sample area. The
cadmium total cross-section and gold (n,$\gamma$) cross-section were
obtained from the ENDF/B-VI data set maintained by the Los Alamos
Nuclear Data Service \cite{lanlt2}.

The absolute value of $k$ for a low efficiency detector may then be
related to the integrated activity through:
\begin{equation}
k={1\over\tilde{A}} \left( \int_0^{1 eV}dE\sigma(E)\Delta_{Cd}
(E)\phi_{th}(E) +\int_{1 eV}^{5.2MeV}dE
\sigma(E)\Delta_{Cd}(E)\phi_{epi}(E) \right)
\end{equation}
We found that the second integral contributes 10\% to the determination
of $k$ for $E_{max}=1$ eV. This second term represents a correction for
the finite absorption of neutrons by Cd at energies above the cut-off
energy (note that this contribution goes to zero if the Cd absorbed no
neutrons above 1 eV). The resulting value of $k$ is $(4.40 \pm
0.88)\times10^{-4}\,\AA^{-1}$ with the uncertainty dominated by that of
the gold foil activity measurements.

The detector was specified to have an ideal efficiency of
$5.47\times10^{-4}\,\AA^{-1}$ (1.91 cm thick, 3 Torr $^3$He with 719
Torr $^4$He as a buffer gas and 38 Torr $N_2$ as a quench gas).
Uncertainties in the measured gas pressures are 0.1 Torr. The absolute
efficiency determined from our gold-foil normalization is consistent
with our specification to the manufacturer and also with an independent
comparison of this detector to a calibrated detector via simultaneous
measurement of the neutron spectrum on the HRMECS instrument at IPNS
\cite{micklich2005,iverson2005}.

\subsubsection{Measured Neutron Energy Spectra}

\begin{figure}
    \begin{center}
    \includegraphics[ ]{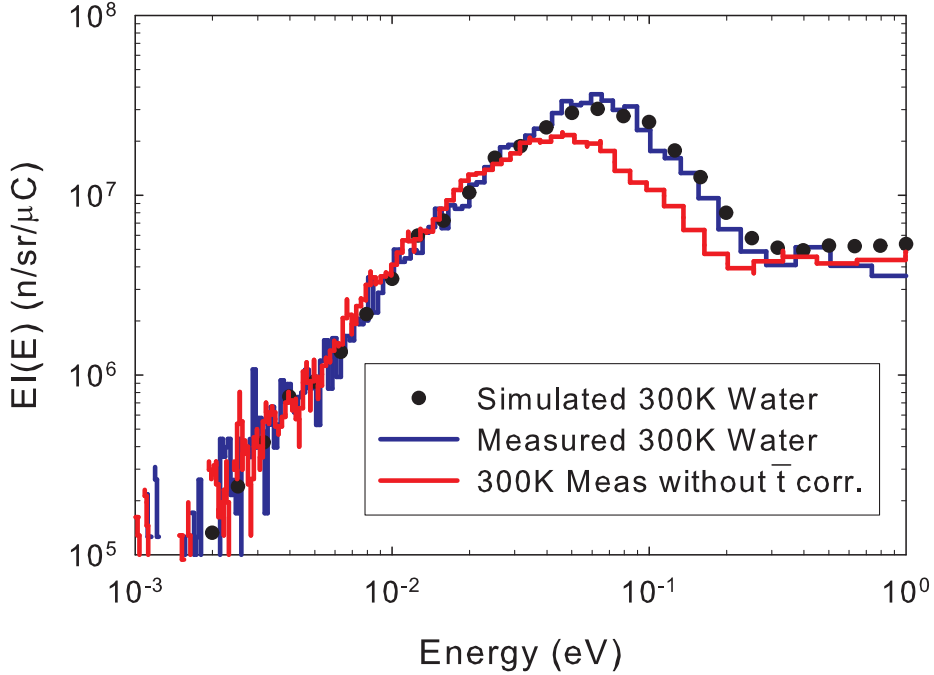}
    \end{center}
    \caption{The water moderated spectral intensity, with and without the $\bar t$
    correction applied to the data compared to simulation.}
    \label{fig:water_flux}
\end{figure}

\begin{figure}
    \begin{center}
    \includegraphics[ ]{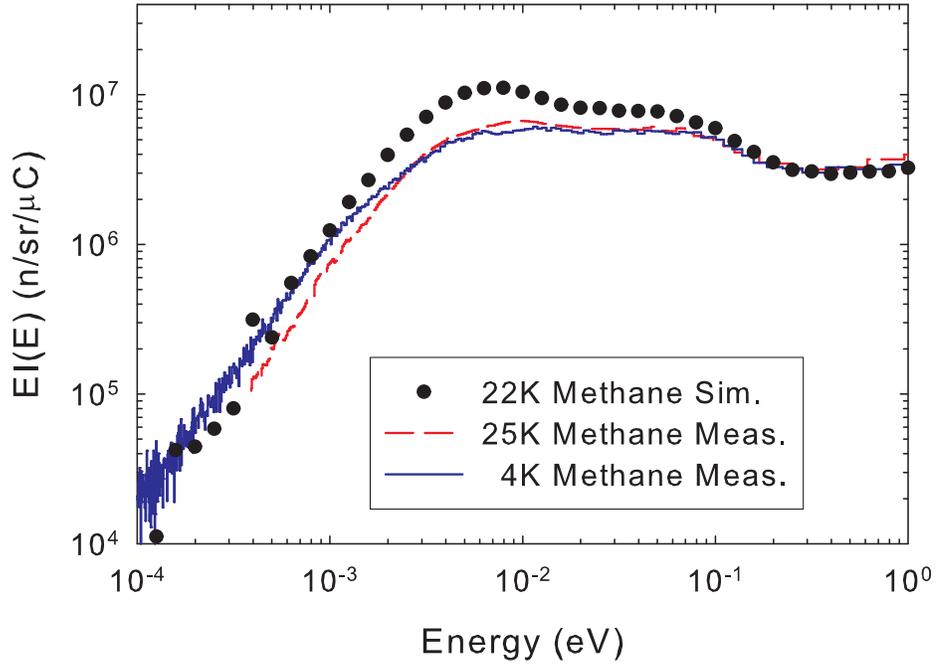}
    \end{center}
    \caption{$EI(E)$ for solid methane at 25 K and  4 K temperatures as a function of neutron
    energy.}
    \label{fig:methane_flux}
\end{figure}

The collimation used to assure that the measurement looked only at
neutrons emitted from the moderator face consisted of two 20 cm long
steel collimators combined with 0.635 cm thick boron nitride (BN) masks
upstream of the first collimator and downstream from the second. The collimator
diameters are 7.62 cm at a distance of 140 cm from the moderator face
and 2.54 cm  at a distance of 570 cm. To facilitate rapid measurement
of spectra at different moderator temperatures we used a high
efficiency (black) LND 25291 neutron detector containing 20 ATM of
$^3$He that was masked by an additional BN aperture of roughly 3 mm
diameter to reduce dead time effects from high instantaneous count
rates at short wavelengths.  To calibrate this detector, the 4 K
spectrum collected with it was fitted to an equivalent spectrum taken
with the thin detector with one free parameter to account for
uncertainty in the precise dimensions of the small aperture in front of
the thick detector. The RMS differences between the neutron spectra
measured by the two detectors between 18 and 1000 meV was only 2.5\%.
The empty moderator and 4 K  methane spectra (shown in figs.
\ref{fig:water_flux} and \ref{fig:methane_flux}) were taken with the
thin detector, and the 25 K spectrum was taken with the black detector
(fig. \ref{fig:methane_flux}).

Tables \ref{tbl:results} and \ref{tbl:results2} show integrated neutron
yields for various  neutron energy groups and the ``1 eV coupling'' for
an empty moderator vessel (water reflector moderated neutrons) and
solid methane moderator. The 1 eV coupling is the value of spectral
intensity (EI(E)) evaluated at 1 eV, and for a 1/E spectrum it is
proportional to the epithermal neutron flux
\cite{carpenter1986,micklich2005}. For this reason it is a common
benchmark used for discussing the performance of a moderator system.
For the empty moderator, the measured integrated neutron yields below
125 meV agree remarkably well with the simulation once the correction
for the emission time delay is applied, as shown in Figure
\ref{fig:water_flux}.  With an empty moderator vessel, the neutron
spectrum depends almost exclusively on the water reflector, for which
the scattering kernels have been well characterized.  The dominant
contribution to the 25\% uncertainty in these measurements is the
absolute efficiency of the detector, with additional contributions
coming from proton current normalization and background determination.

The methane moderator was prepared by liquifying pure methane from the
gaseous state in the moderator vessel. The methane was then cooled from
liquid to the 4 K base temperature in about 2 hours. Cold moderator
neutron energy spectra  were measured  using the high efficiency
neutron detector during the moderator cool-down in order to obtain
spectral information in short exposure times. Once the moderator
reached 4 K, the flux was also measured both with the black detector
and using gold foil activation together with the 1/v detector (as
described in Section \ref{sec:au-det}).

When the moderator is filled with solid methane the measured 1 eV
coupling agrees within the experimental uncertainty with the MCNP
simulations, but agreement with the simulated integrated yields is not
as satisfying (Figure \ref{fig:methane_flux} and Table
\ref{tbl:results2}).  The spectral differences between simulation and
measurement we see below 100 meV may be related to inadequacies in the
existing scattering kernel for methane, or it could be related to
defects in the methane introduced by our method of preparation.
Measurements of solid methane by Grieger et al. have shown that rapid
solidification can introduce inhomogeneities large enough to affect the
total neutron cross-section \cite{grieger1998}. The cross-section is
also known to be sensitive to the total spin state of the identical
protons in the methane molecule, which we did not control in these
measurements. The \textit{smeth22K} kernel is based upon a phonon
expansion with 4 discrete modes to model the excitations of the
hydrogen about the central carbon atom. Two recently-developed
approximate theories exist for the dynamic structure factor of solid
methane which include the contribution from hindered rotors and
correctly account for the spin state of the methane molecules in the
phase II crystal structure that is present below 20 K
\cite{shin2007,Granada2006a,petriw2006}. Comparison of spectra from
moderators condensed under a variety of conditions to predictions from
these recently developed kernels will be the subject of a future
publication.

\begin{table}
\caption{\textit{Empty Moderator Vessel}: Table of experimental and
simulated integral neutron yields from the LENS TMR operated with a  7
MeV proton beam and a 150$\mu$s wide proton pulse. Units of $10^6
n/sr/\mu C$.\label{tbl:results}}
 \label{tbl:results}
\begin{tabular}{|c|c|c|}\hline\hline
\textbf{Energy Group (meV)}&\textbf{MCNP}&\textbf{Meas.}\\\hline
0.20-0.82&$0.0091\pm0.0005$&$<0.04$\\\hline 0.82-3.27&
$0.304\pm0.001$&$0.22\pm0.05$\\\hline
3.27-10&$1.72\pm0.01$&$1.64\pm0.37$\\\hline
10-125&$44.30\pm0.07$&$47.6\pm10.6$ \\\hline 1 eV Coupling &
$5.35\pm0.09$ & $3.8\pm0.9$ \\\hline\hline
\end{tabular}
\end{table}

\begin{table}
\caption{\textit{Solid Methane Moderator}: Table of experimental and
simulated integral neutron yields for 7 MeV proton beam, 150$\mu$s wide
proton pulse. Units of $10^6 n/sr/\mu C$.\label{tbl:results2}}
 \label{tbl:results2}
\begin{tabular}{|c|c|c|c|}
\hline\hline \textbf{Energy Group (meV)}&\textbf{22K MCNP}&\textbf{25K
Meas.}&\textbf{4K Meas.}\\\hline
0.20-0.82&$0.463\pm.002$&$0.18\pm0.04$&$0.44\pm0.09$\\\hline
0.82-3.27&$6.37\pm0.02$&$2.4\pm0.5$ &$3.3\pm0.7$\\\hline
3.27-10&$10.40\pm.01$ &$6.3\pm1.2$ &$6.2\pm1.3$\\\hline
10-125&$19.05\pm0.01$& $14.1\pm2.8$& $13.2\pm2.7$ \\\hline 1 eV
Coupling & $3.25\pm.02$ & $3.5\pm0.7$& $4.0\pm0.8$\\\hline\hline
\end{tabular}
\end{table}

\section{Possible Improvements to Cold Neutron Brightness through Design Modifications}
\label{sec:improvements}
\begin{figure}
    \begin{center}
    \includegraphics[ ]{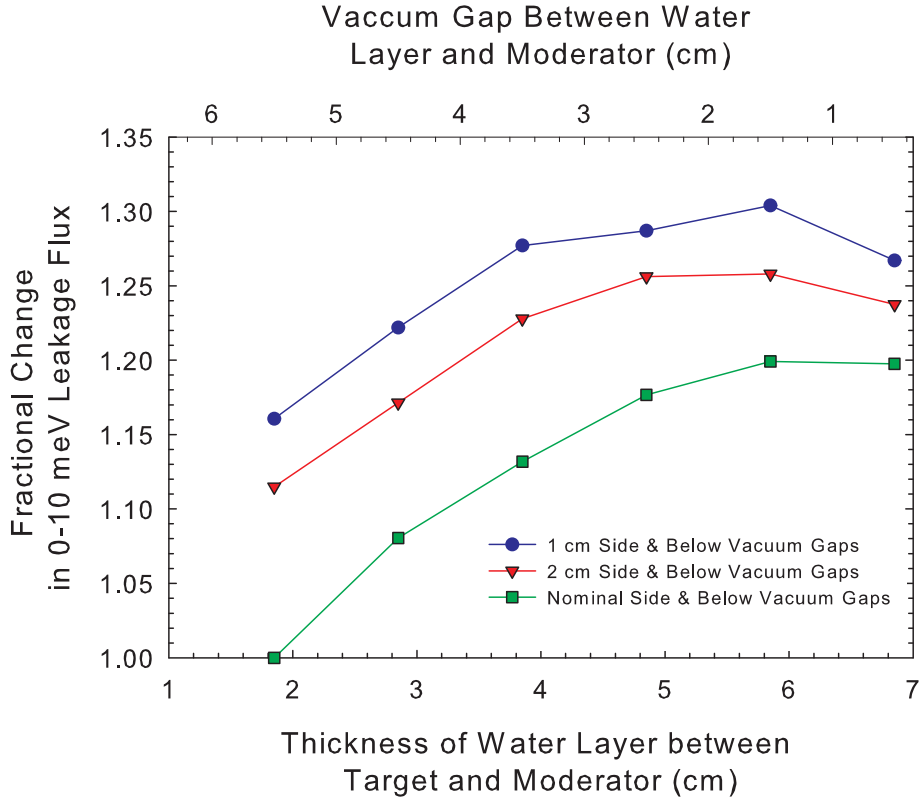}
    \end{center}
    \caption{A simulation of the cold leakage flux enhancement expected from increasing the thickness of reflector material between the target and moderator.  The
    enhancement is greater if voids along the sides and below the
    moderator are reduced to 1 cm from the moderator vessel
    surface.}
    \label{fig:premoderator}
\end{figure}

\begin{figure}
    \begin{center}
    \includegraphics[ ]{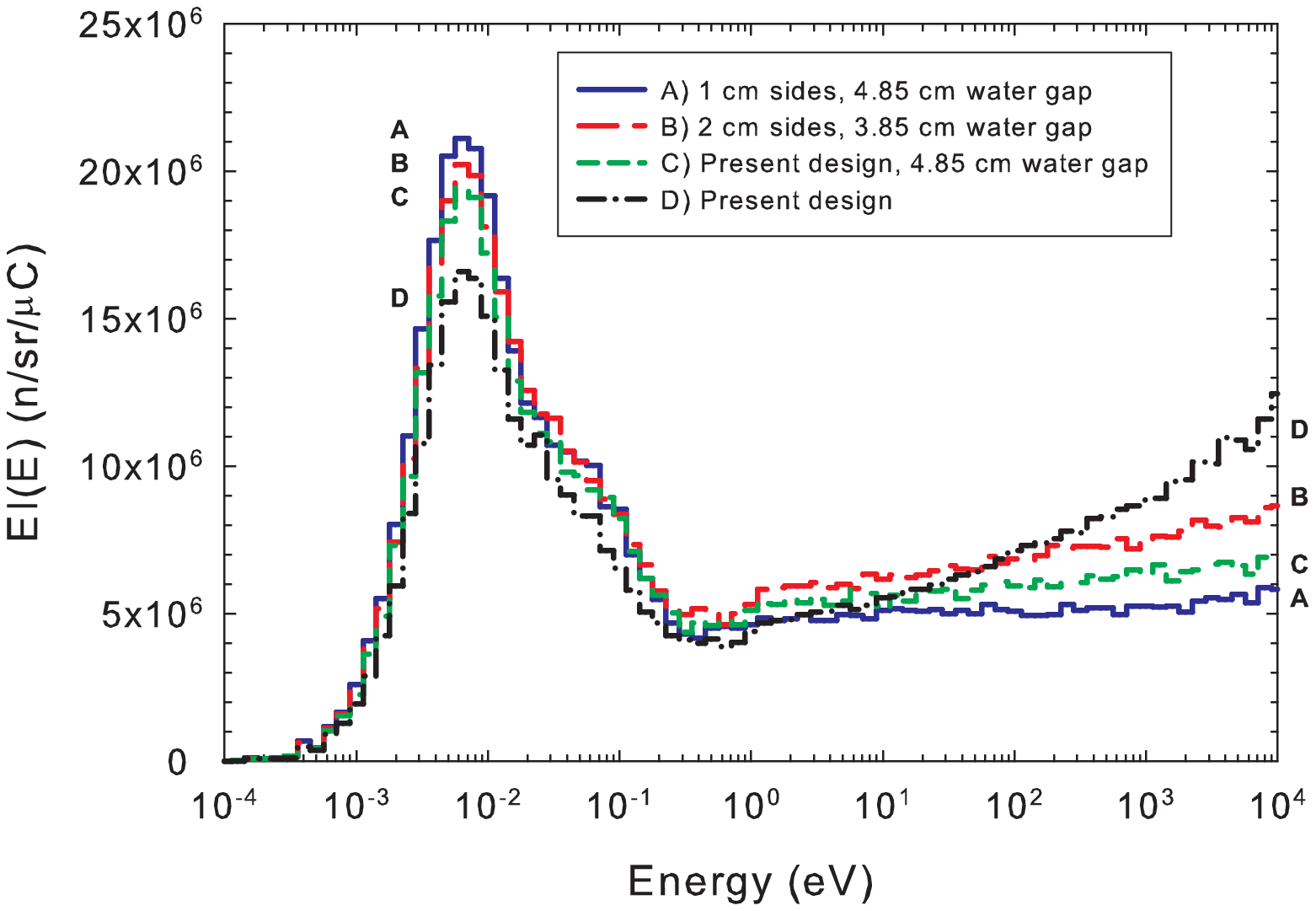}
    \end{center}
    \caption{Simulations of neutron energy spectra for representative configurations
    from Figure \ref{fig:premoderator} show the impact of
    water premoderator layer thickness and vacuum gap
    thickness on cold neutron leakage flux.  The optimal Case
    (A) minimizes vacuum gaps and optimizes premoderator
    thickness.  Case (B) has less cold flux than Case (A) due to enlarged
    side gaps.  Case (C) is similar, where present design's current side gap thickness
    is unchanged and premoderator layer alone is optimized. Case
    (D) is the present design, where the high leakage from the TMR
    causes the slope of the 1-10$^4$ eV flux to increase while at the same time reducing    coupling between the moderator and reflector.}
    \label{fig:gap_spectra}
\end{figure}

There are a number of modifications to the design of the LENS TMR that
could increase the cold neutron brightness from the moderator. The
present design uses an oversized vacuum space around the moderator to
facilitate experiments on a variety of moderator choices during our
initial low-power operation. To investigate the impact of this extra
vacuum space on the neutronic performance, the size of the vacuum gaps
was varied in the MCNP model. The cold leakage flux between 0 and 10
meV from the moderator was simulated as a function of the water layer
thickness between the target and moderator for a number of values for
the vacuum space dimensions below, and to the sides of the moderator.
The results of these simulations are shown in Figures
\ref{fig:premoderator} and \ref{fig:gap_spectra}. Up to a 30\% increase
in the leakage flux below 10 meV is possible simply by reducing the
size of the vacuum gaps. We also notice that decreasing the size of
these vacuum spaces also brings the spectrum from 1 eV to 10 keV closer
to the 1/E behavior expected for an infinite moderator
\cite{carpenter1986}. Figure \ref{fig:gap_spectra} therefore suggests
that some of the gain seen here may come from avoiding under-moderation
in thin sections of the reflector as well as from bringing reflector
material closer to the moderator itself.

Another potential gain for the neutron flux available at LENS may be
obtained by changing to a beryllium reflector. Calculations indicate an
increase in the brightness of an additional 30\% without a significant
increase in the width of the emission time distribution is possible.
The expense and potential for activation due to impurities in the Be
have precluded us from implementing this option in the initial
construction of the facility. The smaller absorption cross section and
less efficient moderation associated with n-Be collisions (compared to
n-H) require a slightly larger Be reflector volume. The present water
reflector was sized with this potential upgrade in mind so it could be
implemented with essentially no impact on the other components of the
TMR. It is also possible that a Be reflector/filter covering the
instrument beam lines could achieve similar gains in flux while
requiring much less Be \cite{Muhrer2005}. In this configuration, cold
Be directly in front of the moderator is used to reflect epithermal
neutrons back while transmitting neutrons with energies below the Bragg
cutoff to the instruments.

Finally, MCNP simulations with a recently developed kernel for solid
methane \cite{shin2007} suggest that increasing the moderator thickness
to 2 cm will increase the cold-neutron brightness and reduce the
spectral temperature. As a first step to check on this possibility we
will investigate the effect of inserting a cold polyethylene
premoderator on the thermal shield in the cryogenics system. This has
the advantage of reducing the thermal and radiation loads on the
moderator (as opposed to the increase in these that would accompany an
increase in the methane thickness itself) while still providing extra
material behind the moderator's front face to cool neutrons below room
temperature. A major goal for the facility over the next several years
is to develop moderator materials and designs suitable for reducing the
spectral temperature of the moderator, and this will be but the first
step in that on-going effort. Enhancements in moderator geometry, such
as grooved and reentrant cavity type moderator will also be explored in
simulation to assess their ability to improve cold neutron yield.

\section{Conclusion}
\label{sec:conclusion}
In this paper we have described the modular design approach taken for
the LENS neutron source and explained the simple physical
considerations which set the scale for the dimensions of the
target/moderator/reflector system and timescales for the time-dependent
neutron field from this strongly-coupled long pulse neutron source. We
have discussed in detail all design features relevant for neutron
production and moderation. We have also outlined the assumptions of our
MCNP model for LENS and compared the measured brightness, energy
spectrum, intensity, and emission time distribution with MCNP
predictions, which agree within the accuracy of the measurements and
the uncertainly of the actual neutronic properties of the moderator.

We hope that this description of the design of LENS will be useful in the future for those readers who choose to construct their own neutron source. The LENS operating regime opens up some interesting possibilities for different types of neutron sources which have not been fully investigated.

\ack{The authors would like to thank P. D. Ferguson and E.B. Iverson for many helpful discussions. We are
also grateful to J.M. Carpenter for the loan of the equipment used at the Intense Pulsed Neutron Source
(IPNS) in reference \cite{carpenter1985} for the emission time measurements. Computational work was supported
by Shared University Research grants from IBM, Inc. to Indiana University and by the National Science
Foundation under Grant No. 0116050. The LENS project is supported by the National Science Foundation (under
grants DMR-0220560 and DMR-0320627), the Indiana 21st Century Fund, Indiana University, and Crane Naval
Surface Warfare Center.

  M. Snow thanks the Institute for Nuclear Theory at the University of Washington for its hospitality and the Department of Energy for partial support during the completion of this work.}

\bibliographystyle{elsart-num}
\bibliography{Lavelle-Main_working.bbl}

\end{document}